\documentclass[prl,superscriptaddress,twocolumn,showpacs,preprintnumbers,amsmath,amssymb]{revtex4-1}

\usepackage{graphicx}
\usepackage{bm}
\usepackage{color}

\def\etal{{\it et~al.}}

\def\s1{\sigma_1(\omega)}

\begin{document}

\preprint{???}

\title{\boldmath Coexistence and competition of magnetism and superconductivity\\on the nanometer scale in underdoped BaFe$_{1.89}$Co$_{0.11}$As$_2$  \unboldmath}

\author{P. Marsik}
\author{K. W. Kim}
\author{A. Dubroka}
\author{M. R\"{o}ssle}
\author{V. K. Malik}
\author{L. Schulz}
\author{C.N. Wang}
 \affiliation{University of Fribourg, Department of Physics and Fribourg Center for Nanomaterials,
Chemin du Mus\'{e}e 3, CH-1700 Fribourg, Switzerland}

\author{Ch. Niedermayer}
\affiliation{Laboratorium for Neutron Scattering, Paul Scherrer Institut \& ETH Z\"{u}rich, CH-5232 Villigen, Switzerland}

\author{A.~J.~Drew}
 \affiliation{University of Fribourg, Department of Physics and Fribourg Center for Nanomaterials,
Chemin du Mus\'{e}e 3, CH-1700 Fribourg, Switzerland}
\affiliation{Queen Mary University of London, Mile End Road, London, E1 4NS, UK}
\author{M.~Willis}
\affiliation{Queen Mary University of London, Mile End Road, London, E1 4NS, UK}

\author{T. Wolf}
\affiliation{Karlsruher Institut f\"{u}r Technologie, Institut f\"{u}r Festk\"{o}rperphysik, D-76021 Karlsruhe, Germany}

\author{C. Bernhard}%
 \email{christian.bernhard@unifr.ch}
\affiliation{University of Fribourg, Department of Physics and Fribourg Center for Nanomaterials,
Chemin du Mus\'{e}e 3, CH-1700 Fribourg, Switzerland}

\date{\today}

\begin{abstract}
We report muon spin rotation ($\mu$SR) and infrared (IR) spectroscopy experiments on underdoped BaFe$_{1.89}$Co$_{0.11}$As$_2$ which show that bulk magnetism and superconductivity (SC) coexist and compete on the nanometer length scale. Our combined data reveal a bulk magnetic order, likely due to an incommensurate spin density wave (SDW), which develops below $T^{mag}\approx32$~K and becomes reduced in magnitude (but not in volume) below $T_c=21.7$~K. A slowly fluctuating precursor of the SDW seems to develop alrady below the structural transition at $T^{s}\approx50$~K. The bulk nature of SC is established by the $\mu$SR data which show a bulk SC vortex lattice and the IR data which reveal that the majority of low-energy states is gapped and participates in the condensate at $T\ll T_c$.

\end{abstract}



\maketitle

The discovery of high temperature superconductivity (HTSC) in the iron arsenide pnictides~\cite{Kamihara08} renewed the interest in the relationship between magnetism and superconductivity (SC). Similar to the cuprate HTSC~\cite{Nieder98,Sanna04,Miller06} and several heavy fermion superconductors~\cite{Pfleid09}, SC emerges here in close proximity to an antiferromagnetic (AF) or spin-density-wave (SDW) state~\cite{Cruz08,Klauss08,Ni08}. Within the so-called underdoped (UD) regime of the phase diagram it was even found that SC sets in already before static magnetism is suppressed~\cite{Ni08,Drew09,Pratt09,Christian09}. These observations have reinforced speculations that the magnetic and SC orders are intimately related and that spin fluctuations may be at the heart of the SC pairing mechanism~\cite{Mazin08}. Nevertheless, it has been questioned whether the magnetic and SC orders truly coexist and interact on the nanometer length scale. For example, it has been argued that they occur in adjacent but different regions of the phase diagram~\cite{Luetkens09} and only coexist in samples that are spatially inhomogeneous with segregated magnetic and SC phases~\cite{Uemura09}. 

Here we present muon spin rotation ($\mu$SR) and infrared (IR) spectroscopy measurements which establish that in UD BaFe$_{1.89}$Co$_{0.11}$As$_2$ the magnetic and SC orders are bulk phenomena which compete for the same electronic states. 

The BaFe$_{1.89}$Co$_{0.11}$As$_2$ single crystals were grown from self-flux in glassy carbon crucibles and their chemical composition was determined by energy dispersive X-ray spectroscopy (EDX) as described in Ref.~\cite{Hardy09}. The resistivity, $\rho(T)$, and the dc magnetization data as shown in Fig. 1 yield a sharp SC transition with a midpoint at $T_c=21.7$~K. The anomalous upturn in $\rho$($T<50$~K) is characteristic of an UD sample with a structural transition at $T^s\approx50$~K and a related magnetic one at ${{T}^{mag}}\le T^s$~\cite{Ni08,Lester09,Rullier09}.

The $\mu$SR experiments on a mosaic of crystals were performed at the GPS instrument at the $\pi$M3 beamline of the Paul Scherrer Institut (PSI) in Villigen, CH, which provides a 100\% spin polarized muon beam. $\mu$SR measures the time evolution of the spin polarization, $P(t)$, of the implanted muon ensemble using the asymmetry, $A(t)$, of the muon decay positrons~\cite{Schenck86}. The technique is well suited to studies of magnetic and SC materials, as it allows a microscopic determination of the internal field distribution and gives direct access to the volume fractions of these phases. The positive muons are implanted into the bulk of the sample stopping at interstitial lattice sites~\cite{Maeter09} within a $100-200$ $\mu$m thick layer. Each muon spin precesses in the local magnetic field, $B_\mu$, with a precession frequency of ${\nu}_{\mu}={\gamma}_{\mu}\cdot{B}_{\mu}/2{\pi}$, where $\gamma_{\mu}=2\pi\cdot135.5$~MHz/T is the muons gyromagnetic ratio. 

For the IR spectroscopy measurements we used freshly cleaved crystals from the same growth batch. Details of the applied reflection and ellipsometry techniques and the data analysis are described in Ref.~\cite{Kim09} which reports corresponding measurements on optimally doped (OP) BaFe$_{1.87}$Co$_{0.13}$As$_2$. Since the probe depth of the IR light exceeds 100 nm, the IR spectra yield the bulk electronic properties. In particular, they enable one to access the fraction of low-energy electronic states that become gapped and participate in the SC condensate~\cite{Basov06,Kim09}.

\begin{figure}
\vspace*{0cm}
\hspace*{-0.3cm}
\includegraphics[width=6.5cm]{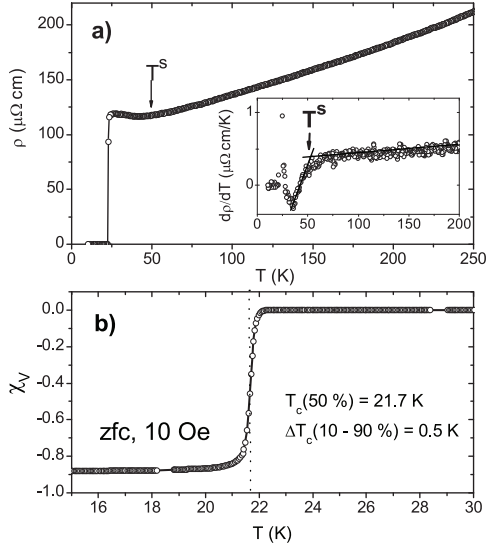}
\vspace{0cm}

\caption{\label{Resistivity}\textbf{(a)} $T$-dependence of the in-plane resistivity, $\rho$, and \textbf{(b)} the zero-field-cooled (zfc) magnetic volume susceptibility, $\chi_{V}$, of BaFe$_{1.89}$Co$_{0.11}$As$_2$ with $H=10$~Oe parallel to the c-axis. Inset: Derivative of $\rho(T)$ showing the change in slope around $T^s\approx50$~K.  
}
\end{figure}

The results of the $\mu$SR experiments are summarized in Fig. 2. The zero-field (ZF) $\mu$SR spectra in Fig. 2(a) establish that static magnetism develops below $\sim40$ K. They have been fitted (solid lines) with the function,\begin{equation} P\left( t \right)=P\left( 0 \right)\cdot \left[ {{A}_{f}}\exp {{\left(-\lambda _{f}^{ZF}t \right)}^{\beta }}+{{A}_{s}}\exp \left(-\lambda _{s}^{ZF}t \right) \right],
\end{equation} where the two terms account for the fast (f) and slowly (s) depolarizing parts of the signal, respectively. The rapid depolarization of P(t) at low $T$ without a trace of an oscillation suggests that the magnetic state is either incommensurate or strongly disordered. The IR data shown below support an incommensurate spin density wave (SDW). The low-$T$ value of ${{\lambda }_{f}}^{ZF}\approx \text{8}$ ${\mu}$s$^{-1}$ yields an estimate of the average local field of $\left\langle B_{\mu }^{ZF} \right\rangle\approx {{\Lambda }^{ZF}}/{{\gamma }_{\mu }}\approx 10$~mT which is around 5\,\% of the value in undoped BaFe$_2$As$_2$ \cite{Klauss08,Aczel08}. The amplitude of $A_f\approx0.7$ confirms that the magnetic volume fraction is at least 70\,\%. However, it is likely to be even higher since $\vec{B}_{\mu }^{ZF}$ is not likely to be orthogonal to $\vec{P}\left( t=0 \right)$. 

This magnetic volume fraction has been more accurately determined from the transverse-field (TF) $\mu$SR data as shown in Fig. 2(b) where the direction of the external field of $H_{ext}=300$~mT is well defined. The TF-$\mu$µSR spectra were fitted with the function:\begin{equation}
\begin{split}
P\left( t \right)= P\left( 0 \right)\cdot & \left[{{A}_{f}}\cdot \cos \left( \gamma B_{{\mu},f}t \right)\cdot \exp \left(-\lambda _{f}^{TF}t \right)+ \right. \\
  & \left. +{{A}_{s}}\cdot \cos \left( \gamma {{B}_{\mu ,s}}t \right)\cdot \exp \left(-{{\lambda }_{s}}t \right) \right].
\end{split}
\end{equation} At 100 K there is only a weak depolarization of the oscillatory signal which typically arises from nuclear moments. At 25 K and 5 K the major part of the signal depolarizes very rapidly due to the presence of static electronic moments. At both temperatures we obtain the same amplitudes of $A_f\approx0.92$ and $A_s\approx0.08$ that are characteristic of a bulk magnetic state. The latter is in fact well accounted for by the background of muons that stop in the sample holder or the cryostat shields and windows. Our TF-$\mu$SR data therefore provide compelling evidence for a magnetic phase that occupies more than 90\% (likely even 100\%) of the volume. This bulk magnetic state may be strongly disordered and exhibit a large spatial variation of the magnitude of the moments. Nevertheless, since the magnetic stray fields away from regions with AF or SDW order are known to rapidly fall off on the scale of a few nanometers, any non-magnetic regions must be extremely small and well intermixed with the magnetic ones.
	
\begin{figure}
\vspace*{0cm}
\hspace*{-0.2cm}
\includegraphics[width=8.8cm]{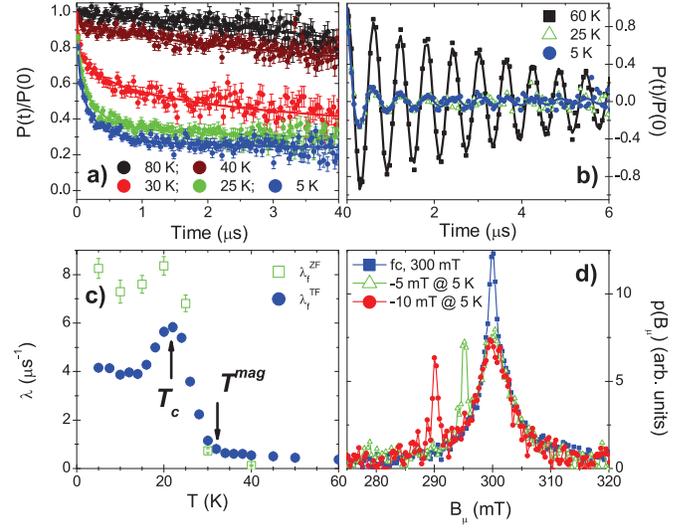}
\vspace{-0.5cm}

\caption{\label{muSR}\textbf{(a)} ZF-$\mu$SR time spectra (symbols) showing a rapid depolarization below about 40~K. Solid lines show fits with eq. (1). \textbf{(b)} TF-$\mu$SR spectra (symbols) at $H_{ext}=300$~mT plotted in a rotating reference frame ($\omega_{RRF}=39.5$~MHz). Solid lines show fits with eq. (2). \textbf{(c)} $T$-dependence of the depolarization rates, $\lambda _{f}^{TF}$ and $\lambda _{f}^{ZF}$. \textbf{(d)} TF-$\mu$SR lineshapes showing the distribution of local magnetic fields, $p\left( B_{\mu}\right)$, during a so-called pinning experiment as described in the text.
}
\end{figure}
	
\begin{figure}
\vspace*{0cm}
\hspace*{-0.5cm}
\includegraphics[width=7cm]{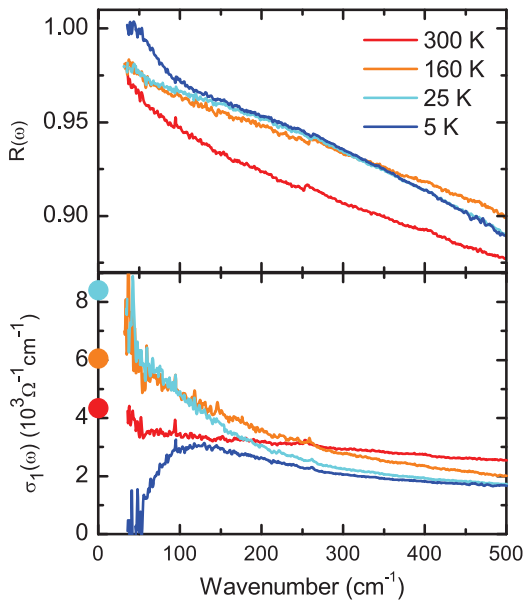}
\vspace{0cm}

\caption{\label{Reflectivity}$T$-dependent spectra of \textbf{(a)} the FIR reflectivity, $R\left(\omega\right)$, and \textbf{(b)} the in-plane optical conductivity, $\sigma_1\left(\omega\right)$, of BaFe$_{1.89}$Co$_{0.11}$As$_2$ with $T_c=21.7$~K. Solid circles show $\sigma^{dc}$ as obtained from Fig. 1(a).}
\end{figure}

The $T$-dependence of the magnetic order is detailed in Fig. 2(c) in terms of the relaxation rates $\lambda_f^{ZF}$ and $\lambda_f^{TF}$. The transition to the static magnetic state at $T^{mag}\approx32$~K is clearly identified from the steep increase of $\lambda_f^{TF}$. There is also a weak, yet noticeable increase of $\lambda_f^{TF}$ above $T^{mag}$ which is likely due to a fluctuating magnetic state as discussed below in the context of the IR data. The most remarkable feature is the anomaly at $T_c=21.7$~K below which $\lambda_f^{TF}$ decreases again. This observation agrees with a previous neutron diffraction study which revealed a similar SC-induced reduction of the intensity of a magnetic Bragg peak~\cite{Pratt09,Christian09}. The neutron data exclude a reorientation of the magnetic moments or a change in their periodicity. However, they cannot distinguish between a SC-induced reduction of the order parameter amplitude and a decrease of the magnetic volume fraction~\cite{Christian09}. Our $\mu$SR data provide this additional information and thus help to identify the scenario of a SC-induced reduction of the bulk magnetic order parameter.

Our $\mu$SR data also confirm the bulk nature of SC. Figure 2(d) displays a so-called pinning experiment which demonstrates that a bulk SC vortex lattice develops below $T_c$~\cite{Sonier94,Miller06}. It shows so-called $\mu$SR-lineshapes, as obtained from a fast-Fourier transformation of the TF-$\mu$SR time spectra, which detail the distribution of the magnetic field probed by the muons, $p\left({B_\mu}\right)$. The first measurement (blue squares) has been performed directly after field-cooling (fc) the sample to 5 K at $H_{ext}=300$~mT. Before the second measurement (green triangles) $H_{ext}$ was reduced by 5 mT to 295 mT. Finally, $H_{ext}$ was reduced by another 5 mT to 290 mT before the third measurement (red circles). The lineshapes consist of a weak, narrow peak which corresponds to the background muons that stop outside the sample and a dominant, broad peak which arises from the muons that stop inside the magnetic sample. The weight of these peaks agrees with $A_f\approx0.92$ and $A_s\approx0.08$ as obtained from the time spectra.
The striking result is that the broad part of the $\mu$SR lineshape remains fully pinned, while only the narrow peak is following the changes of $H_{ext}$. This behavior highlights a complete pinning of the magnetic flux density in the sample which is the hallmark of a bulk type-II superconductor. While the pinning centers are regions where SC is reduced or even vanishing, for the SC vortices to be rigid and the pinning to be effective, the majority of the volume must be SC. The circumstance that the entire sample is at the same time in a magnetic state, is manifested by the broadening of the $\mu$SR lineshape which is much larger than expected for a SC vortex lattice~\cite{Luetkens08,Drew08,Bernhard09}. 
	
Our IR data in Fig. 3 confirm the bulk nature of SC. Similarly as previously reported for OP samples~\cite{Kim09,Heumen09,Li08}, they reveal characteristic SC-induced changes due to the formation of a bulk SC state with multiple and nearly isotropic BCS-like energy gaps~\cite{Kim09}. Notably, the reflectivity, $R\left( T\ll T_c \right)$, approaches unity around 50~cm$^{-1}$ and the real part of the optical conductivity, ${{\sigma }_{1}}\left( T\ll{{T}_{c}} \right)$, has a pronounced lower gap edge around 50~cm$^{-1}$ below which it vanishes within the accuracy of the data. This highlights that the vast majority of the low-energy electronic states is captured by the SC energy gap(s) and participates in the SC condensate. The so-called Ferrell-Glover-Tinkham sum-rule requires that the spectral weight, $SW\left( \Omega  \right)=\frac{120}{\pi }\int_{0}^{\Omega }{{{\sigma }_{1}}\left( \omega  \right)\,d\omega }$, with $\sigma_1\left(\omega\right)$ in units of $\Omega^{-1}$cm$^{-1}$,  remains unaffected by the SC transition for $\Omega \ge 6-10\,\Delta^{SC}$~\cite{Tinkhambook}. Accordingly, the so-called missing SW due to the gap-like suppression of the regular part of $\sigma_1$ is transferred to a $\delta$-function at the origin which accounts for the loss-free response of the SC condensate. Its weight, $S{{W}^{\delta }}=4\pi {{e}^{2}}\cdot \frac{{{n}_{s}}}{m^*}=\frac{{{c}^{2}}}{\lambda _{ab}^{2}}$, with the condensate density, $n_s$, effective mass, $m^*$, and magnetic penetration depth, $\lambda_{ab}$ , can be deduced either from the missing SW in $\sigma_1$, or likewise from the inductive response in the imaginary part with $\sigma _{2}^{\delta }\left( \omega  \right)=\frac{{{c}^{2}}}{\omega \cdot \lambda _{ab}^{2}}$~\cite{Kim09,Basov06}. For our UD sample we obtain $\lambda_{ab}\approx305(10)$~nm corresponding to a value of $n_s$ that is reduced by about 20\,\% as compared to OP BaFe$_{1.87}$Co$_{0.13}$As$_2$ with $\lambda_{ab}\approx270$~nm~\cite{Kim09}.  

\begin{figure}
\vspace*{0cm}
\hspace*{0cm}
\includegraphics[width=8.2cm]{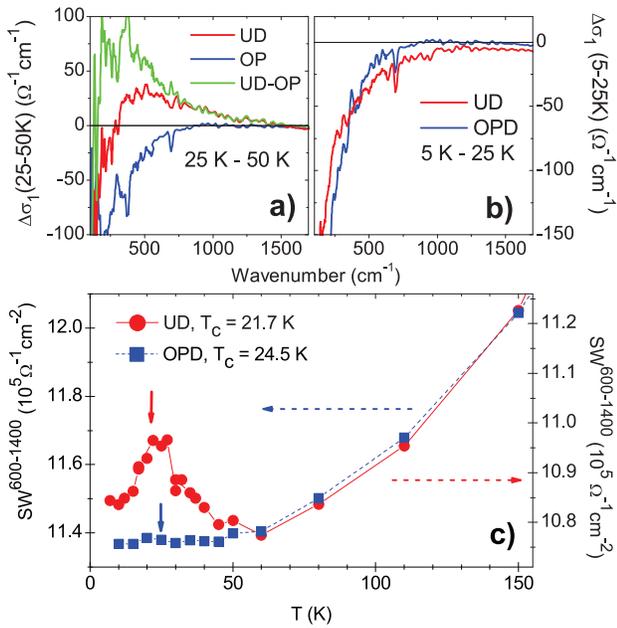}
\vspace{0cm}

\caption{\label{Ellipsometry}Comparison of the $T$-dependent spectral changes for UD BaFe$_{1.89}$Co$_{0.11}$As$_2$ and OP BaFe$_{1.87}$Co$_{0.13}$As$_2$ as obtained with IR ellipsometry. \textbf{(a)} and \textbf{(b)} Difference spectra of the optical conductivity, $\sigma_1$, between 25~K and 50~K and between 5~K and 25~K, respectively. \textbf{(c)} $T$-dependence of the spectral weight integrated from 600 to 1400 cm$^{-1}$. $T_c$ is marked by the solid arrows.
}
\end{figure}

Figure 4 shows that our optical spectra also contain the signatures of the SDW state and its competition with SC. Similar to other SDW systems~\cite{Dressel book}, including undoped  BaFe$_2$As$_2$~\cite{Hu08}, a part of the low-energy SW is transferred to higher energy where it gives rise to a pair breaking peak (PBP) near the gap edge at $2\Delta^{SDW}$.  This PBP shows up in Fig. 4(a) as a broad maximum in $\Delta {{\sigma }_{1}}={{\sigma }_{1}}$(25~K)$-{{\sigma }_{1}}$(50~K) for the UD sample (red line). For the OP sample (blue line) it is absent and $\Delta\sigma_1$ only reveals the signatures of a narrowing of the Drude-response. Assuming that in the UD sample the Drude-response is subject to a similar narrowing, we subtracted this contribution to single out the changes due to the SDW (green line). This reveals a broad maximum corresponding to $2\Delta^{SDW}\approx200-300$~ cm$^{-1}$ which, similar to $T^{mag}\approx32$~K and $T^s\approx50$~K, is reduced by $\sim$3-4 times as compared to undoped BaFe$_2$As$_2$ with $2\Delta^{SDW}\approx900$~cm$^{-1}$ and  $T^{mag}=T^{s}=138$~K~\cite{Hu08}. In agreement with the magnetic order parameter as deduced from $\mu$SR, we obtain a SW of the broad maximum due to the PBP of $\sim$25000~$\Omega^{-1}$cm$^{-2}$ which is around 5\,\% of the value in BaFe$_2$As$_2$~\cite{Hu08}. The $T$-dependence of the PBP is detailed in Fig. 4(c) which compares the evolution of the SW between 600 and 1400 cm$^{-1}$, $SW^{600-1400}$, in the UD and OP samples. In the latter,  $SW^{600-1400}$ decreases monotonically towards low $T$ with no anomaly at $T_c$. For the UD sample, clear deviations occur at low $T$ which signify that the PBP gradually develops below $T^s\approx50$~K as defined by the anomaly of  $\rho(T)$ in Fig. 1. There is some extra increase below $T^{mag}\approx32$~K, but this is rather small as compared to the one of $\lambda^{TF}$ in Fig. 2(c). This discrepancy may be resolved in terms of a precursor SDW which develops already below $T^s$ as predicted e.g. in  Ref.~\cite{Mazin09}. Figure 2(c) shows indeed a weak, yet significant increase of $\lambda^{TF}$ below about 50~K which is compatible with a precursor SDW that fluctuates on the microsecond time scale (but is quasi-static on the sub-picosecond scale of the IR experiment). 
Finally, the decrease of the PB peak below $T_c=21.7$~K resembles the one of $\lambda^{TF}$ in Fig. 2(c) and confirms that the SC and SDW orders are competing for the same low-energy electronic states. We remark that this reduction of $SW^{600-1400}$ cannot be directly caused by the formation of the SC energy gap. This is evident from Fig. 4(b) where $\Delta {{\sigma }_{1}}={{\sigma }_{1}}$(5~K)$-{{\sigma }_{1}}$(25~K) for the OP sample (where $\Delta^{SC}$ is even slightly larger than for the UD one~\cite{Kim09}) essentially vanishes above 600 cm$^{-1}$. As compared to the SW of the SC condensate of $SW^{\delta}\approx7*10^5~\Omega^{-1}$cm$^{-2}$, the SW of the SDW thus amount to only $\sim$3-4\,\% above $T_c$ and $<$2\,\% at $T\ll T_c$. The SDW state, while it occupies the entire sample volume (as shown by $\mu$SR), thus involves only a very small fraction of the electronic states in the vicinity of the Fermi-level. 
These trends are indeed consistent with the scenario of spin fluctation mediated SC where the competing SDW and SC orders are separated in momentum space~\cite{Vorontsov09}. The SDW develops here only in those parts of the Fermi-surface where the nesting condition overcomes a critical threshold. These regions are shrinking as a function of Co-doping and the SC pairing, which is determined by the averaged spin susceptibility, becomes strongest just as the SDW finally dissapears.

In summary, we performed $\mu$SR and IR experiments on weakly underdoped BaFe$_{1.89}$Co$_{0.11}$As$_2$ which establish that superconductivity and static magnetism coexist and compete on the nanometer length scale. In particular, our combined data establish the bulk nature of the superconducting and magnetic orders and show that these compete for the same electronic states. The magnetic order below $T^{mag}\approx32$~K is likely due to an incommensurate SDW of which a fluctuating precursor develops already below the structural transition at $T^{s}\approx50$~K. While this SDW occupies the entire sample volume, it involves only a small fraction of the electronic states.

\begin{acknowledgments}
This work was partially performed at the IR-beamline of the ANKA synchrotron at FZ Karlsruhe, D, and at the Swiss muon source, S$\mu$S, at Paul Scherrer Institut, Villigen, CH. It was financially supported by the Schweizerischer Nationalfonds (SNF) via grants 200020-119784 and 200020-129484, the NCCR-MaNEP and the Sciex-NMS$^{CH}$ fellowship No. CZ0908003. AJD acknowledges financial support from the Leverhulme Trust.
\end{acknowledgments}

\end{document}